\documentclass[superscriptaddress,prl,twocolumn]{revtex4}  
\usepackage{graphicx}

\begin{document}

\title{Statistical physics of the Schelling model of segregation}

\author{Luca Dall'Asta} 
\affiliation{Abdus Salam International Center for Theoretical Physics, Strada
  Costiera 11, 34014, Trieste (Italy)} 
\author{Claudio Castellano}
\affiliation{SMC, INFM-CNR and Dipartimento di Fisica, ``Sapienza'' Universit\`a di Roma, P.le A. Moro 2, 00185 Roma, (Italy)}
\author{Matteo Marsili} 
\affiliation{Abdus Salam International Center for Theoretical Physics, Strada
  Costiera 11, 34014, Trieste (Italy)} 


\begin{abstract}
We investigate the static and dynamic properties of a celebrated model of
social segregation, providing a complete explanation of the mechanisms leading to segregation both in one- and two-dimensional systems. 
Standard statistical physics methods shed light on the rich phenomenology of this simple model, exhibiting static phase transitions typical of kinetic constrained models, nontrivial coarsening like in driven-particle systems and percolation-related phenomena.
\end{abstract}

\maketitle

Individuals with similar ideas, habits or preferences tend to create
cliques and cluster in communities, which results in segregation at
the scale of the society.
A first theoretical analysis of segregation phenomena in social
environments was performed by T.C. Schelling, who defined a model in
which agents divided in two species move on a checkerboard according
to a given utility function \cite{schelling}. Within this simple
model, Schelling was able to show that segregation occurs even when
individuals have a very mild preference for neighbors of their own
type, as long as they are allowed to move in order to satisfy their
preference.\\
Interest for Schelling's results has grown recently among social scientists \cite{vriend,kirman}, mathematicians \cite{math} and statistical physicists
\cite{stauffer}. In particular, Ref.~\cite{kirman} suggested relations with models of binary mixtures in physics. Indeed,
the way in which segregation takes place in the Schelling model
resembles the coarsening processes governing phase separation kinetics
\cite{gunton,bray-rev}.  However, apart from a qualitative picture,
Ref.~\cite{kirman} does not provide any
quantitative measure of coarsening from which the scaling behavior of
the segregation process could be inferred.  On the other hand, looking
at the non-equilibrium process of such a model, several questions
spontaneously arise: Does the system fully segregate?  What are the
properties of the stationary state?  If coarsening takes place in this
system, does it fall in one of the known universality classes?

In this Letter, we shed light on the static and dynamical properties
of the Schelling model in one and two dimensions, applying methods
taken from different fields of statistical physics, such as (vacancy
mediated) coarsening dynamics \cite{bray-rev},
diffusion-annihilation particle systems
\cite{racz}, and kinetic constrained systems \cite{ritort,desmedt}.\\
The model is the same investigated in Ref.~\cite{kirman}: individuals
are initially distributed at random on a line ($d=1$) or a square
lattice ($d=2$) of $N=L^d$ sites.  The occupation of each site $i$ is
described by a spin variable $\sigma_{i}$, taking values $\sigma_i=0$
if the site is empty and $\sigma_i=\pm 1$ if the site is occupied by
an individual of type $\pm 1$.
Let $\rho_{0}=N_{0}/N$ and  $\rho_{\pm}=N_{\pm}/N$ be the densities of vacancies and of occupied sites, of either type, respectively.\\
The general principle governing the dynamics is that an individual can
tolerate at most a fraction $f=f^{*}$ of his (Moore) neighbors to be
different from him.  If $f \leq f^{*}$ the individual is {\em happy}
or has utility $1$ otherwise he feels {\em unhappy} (utility $0$).
Unless otherwise stated we consider $f^{*}=1/2$, as in the original
Schelling model.  Two types of dynamics are possible: {\em
  constrained} dynamics (called ``solid model'' in Ref.
\cite{kirman}), where only unhappy individuals are allowed to move, as
long as they are able to find a vacancy where they can be happy; and
{\em unconstrained} dynamics (called ``liquid model'' in
\cite{kirman}) where agents are allowed to move to vacancies as long
as their situation does not get worse off.  In both cases, following
Ref.~\cite{kirman}, we assume an infinite range dynamics, i.e.
individuals can move to any suitable vacancy, irrespective of
distance. This spin-vacancy exchange dynamics conserves only globally
the magnetization, but not locally.  Natural quantities to
characterize the state of the system are the densities of unhappy
sites $u(t)$ and of interfaces $n(t)$, defined as the fraction
of neighboring spins of opposite type.\\
We stress that, while it might be appealing to introduce a notion of energy, as e.g. the number of unhappy individuals, this might be confusing as the dynamics is not derived from an energy functional, as in physics. Individuals move solely in order to maximize their utility, with no regard to the welfare of the fellow neighbors. For example, even if a displacement is beneficial to the mover, this might make some of his neighbors unhappy with the composition of their new neighborhood. So a move may cause an {\em increase} of the number of unhappy individuals.\\
In the following, we focus on the static properties of the constrained model and the dynamical properties of the unconstrained one, showing that statistical physics allows one to understand many aspects of the rich phenomenology of the model. We shall use the $1d$ case to uncover the main properties of the segregation process and show that this provides key insight on the behavior of the 2d case.

{\em The constrained model -} Let us first discuss the case where only
``condensation'' moves, which increase the utility of the moving
agent, are allowed. At each time step, an individual is drawn at
random and swapped with a
randomly chosen vacancy if his utility increases. If there is no such
vacancy, the agent is not displaced. This dynamics converges in a finite time to
frozen states, called (Myopic) Nash Equilibria (NE), where no agent
can find a better location in terms of utility given the location
chosen by the others \cite{vriend,math}. Segregation occurs up to a finite length, both in $d=1$ and $2$. \\
In $1d$, it is possible to investigate the blocked configurations by means of a
static approach \cite{edwards}, that consists in considering the
ensemble of such configurations {\em with equal a priori weight}.  
The {\em partition function} for this ensemble is
\begin{equation}
\label{Z}
\mathcal{Z}_L(h,\mu,\beta)=\sum_{\sigma\in \mathcal{C}}e^{hM(\sigma)+\mu N(\sigma)+\beta U(\sigma)}
\end{equation}
where $\mathcal{C}$ is the set of all blocked configurations $\sigma=(\sigma_1,\ldots,\sigma_L)$ of size $L$, with $\sigma_i=0,\pm $ (with periodic boundary conditions), $M(\sigma)=\sum_i\sigma_i$ is the magnetization, $N(\sigma)=\sum_i\sigma_i^2$ is the number of individuals and $U(\sigma)$ is the number of unhappy individuals.
The coefficient of the term proportional to $v^{L(\rho_+-\rho_-)}w^{L(\rho_++\rho_-)}$ in $\mathcal{Z}_L(\ln v,\ln w,0)$ is the number of NE for a chain with $L\rho_+$ individuals of one type and $L\rho_-$ individuals of the other type.
\begin{figure}
\centerline{
\includegraphics*[width=0.4\textwidth]{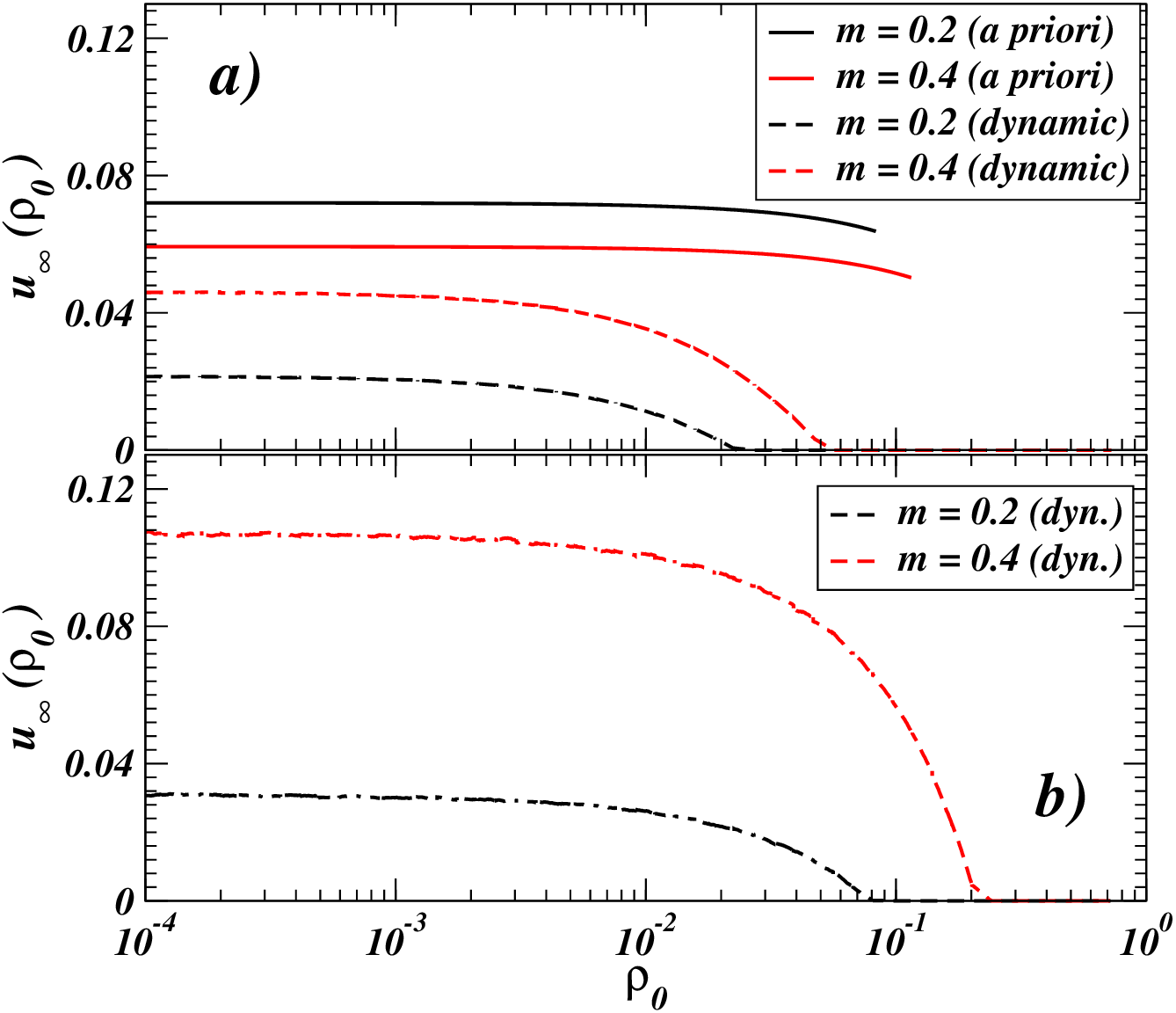}}
\caption{a) Density of unhappy spins $u_{\infty}$ in the blocked configurations of the constrained $1d$ model as a function of the density of vacancies $\rho_{0}$ and for different values of magnetization $m>0$. Solid lines are obtained with the transfer matrix approach for the case of equal a priori weight; the dashed lines are the results of the dynamical computation.  b)  The quantity $u_{\infty}(\rho_{0})$ computed dynamically for a $2d$ system of size $L=128$.
} 
\label{fig1}
\end{figure}
$\mathcal{Z}_L$ can be computed using the transfer matrix technique, following the same approach of Ref.~\cite{desmedt}. In brief, one can write $\mathcal{Z}_L={\rm Tr}\,\mathcal{T}^L$, where the transfer matrix $\mathcal{T}$ relates the statistical weights of configurations of length $\ell+1$ to those of configurations of length $\ell$. In order to build configurations without constraints, $\mathcal{T}$ must have elements $T^{(\sigma_{\ell-1},\sigma_{\ell}),(\sigma_{\ell},\sigma_{\ell+1})}$ which depend on three consecutive spins, i.e. $\mathcal{T}$ is a $9\times 9$ matrix \footnote{Clearly $T^{(p,q),(r,s)}=0$ if $r\neq q$ or if spin $r=q$ in the triplet $\ldots,p,r,s,\ldots$ cannot appear in a NE configuration. For example, $T^{(-,+),(+,0)}$ contains an unhappy $+$ site and it has to vanish if $\mathcal{T}$ is to generate NE without unhappy individuals of type $+$.}.\\
It is important to realize that NE can be of three types only. Either a configuration is blocked because there are no unhappy sites, or because there are unhappy sites with, say, $\sigma_i=+$ but there are no vacancies which could accommodate them (such as e.g. $\ldots+0+\ldots$). This implies that the partition function has the form
\begin{equation}
\label{Z3}
\mathcal{Z}_L(h,\mu,\beta)={\rm Tr}\,\mathcal{T}^L_0+{\rm Tr}\,\mathcal{T}^L_++{\rm Tr}\,\mathcal{T}^L_-
\end{equation}
where $\mathcal{T}_0$ generates all configurations with no unhappy sites, while  $\mathcal{T}_+$ and $\mathcal{T}_-$ those with unhappy sites of type $+$ or $-$, respectively, but no corresponding vacancies.\\
For $L\gg 1$, each term in Eq. (\ref{Z3}) is dominated by the largest eigenvalue $\lambda_q(h,\mu,\beta)$ of the corresponding matrix $\mathcal{T}_q$ ($q=0,\pm$). Hence $\log\mathcal{Z}_L/L\cong \max_q \log \lambda_q(h,\mu,\beta)$. The number of NE with $m=\rho_+-\rho_-$ and a density $\rho_0$ of vacancies is given by $e^{LS(m,\rho_0;\beta=0)}$ where the entropy $S(m,\rho_0 ;\beta)=\max_{h,\mu,q}$ $\left[  \log \lambda_q(h,\mu ,\beta)-hm-\mu(1-\rho_0)\right]$
is obtained from $\lambda_q$ via Legendre  transform. This allows us to access the statistical properties of NE depending on the various parameters of the problem. \\
In particular, it is interesting to look at the behavior of the static density $u_\infty=\partial_{\beta} S(\beta=0)$ of unhappy individuals as a function of $\rho_{0}$.
For $\rho_0>\rho_0^*(m)$ the solution is dominated by the configurations with no unhappy site ($u_\infty=0$, $\lambda_0>\lambda_\pm$) but the converse is true for small density of vacancies [$\rho_0<\rho_0^*(m)$]. Indeed the density $u_\infty(\rho_0)$ of unhappy sites features a jump at $\rho_0^*$ (Fig.~\ref{fig1}a). \\
Such first order phase transition  is typical of cases where the partition sum can be separated in different components as in Eq. (\ref{Z3}). For $m> 0$ the leading contribution to $\mathcal{Z}_L$ comes from the term $\mathcal{T}^L_+$, implying that NE are characterized by a finite density of majority type individuals who cannot find a suitable location. For $m=0$, the symmetry between the two types of agents is spontaneously broken for $\rho_0<\rho_0^*(0)$, in the sense that a randomly drawn NE will typically have a fraction $u_\infty>0$ of unhappy individuals of one type, all individuals of the other type being happy.

The results of the static approach qualitatively agree with simulations, where blocked configurations are selected by the dynamics, starting from a random initial condition (see Fig. \ref{fig1}a).
When we start from $m=\rho_+-\rho_-\neq 0$, we again observe a non-zero fraction of unhappy individuals ($u_\infty>0$) of the majority type, for small enough $\rho_0$.
At odds with the static computation, {\em i)} the transition is continuous rather than discontinuous and {\em ii)} when $m=0$ no spontaneous symmetry breaking is observed. \cite{preparation} \footnote{For $m=0$ and $\rho_0$ small the stationary number of unhappy sites scales as $U=\sqrt{L}f(\sqrt{L}\rho_0)$, with $f(x) =0$ for $x>x_c$. The reason might be that the initial difference $\Delta U$ in the number of unhappy individuals of the two types is $\sim \sqrt{L}$ for $m=0$ (whereas it is $\sim L$ for $m\neq 0$) and that $\Delta U$ typically does not increase in the dynamics.}  

The same qualitative picture applies to the original two dimensional case, with Moore neighborhood. In $d=2$,  the transfer matrix approach to the static problem becomes unfeasible. Progress can be done by MonteCarlo simulations of an appropriately defined auxiliary model \cite{preparation}. This shows the presence of a discontinuous transition in $u_\infty$ as a function of $\rho_0$, similar to the one occurring in $1d$ (not shown). Moreover, Fig.~\ref{fig1}b shows that  the $u_\infty(\rho_{0})$ curves produced by the dynamics of Schelling's model have the same qualitative behavior of the $1d$ case.

\begin{figure}[t]
\centerline{
\includegraphics*[width=0.4\textwidth]{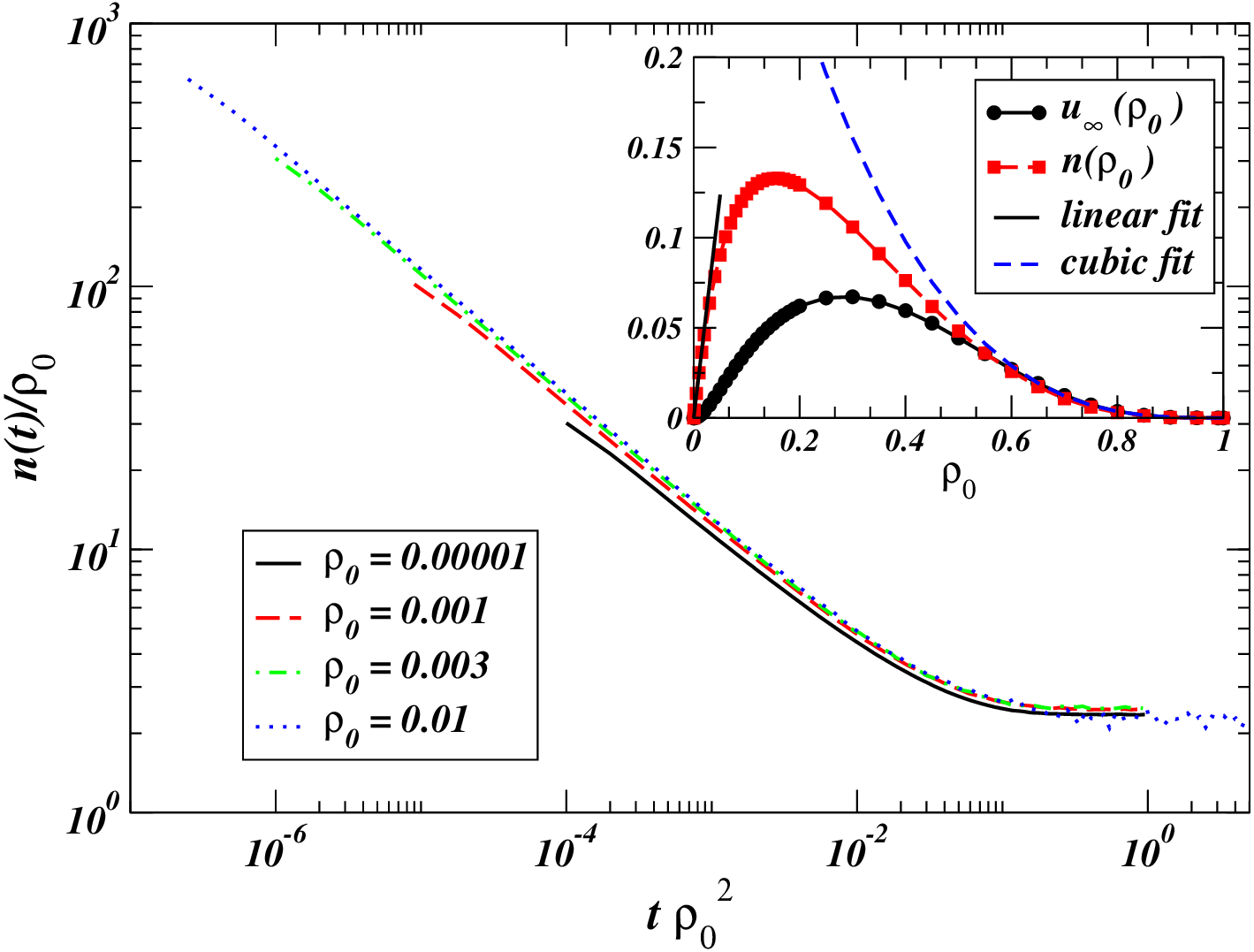}}
\caption{Scaling behavior of the density of interfaces $n(t)$ observed in the limit of low density of vacancies $\rho_{0}$ for a $1d$ system of size $L=10^6$ and zero magnetization $m=0$. The inset shows the behavior of the stationary value of the densities of unhappy spins and interfaces as a function of $\rho_0$.    
} 
\label{fig2}
\end{figure}

{\em In the unconstrained model}, 
at each time step a randomly chosen agent is relocated into a randomly chosen vacancy (long-range diffusion) when his utility does not decrease.
No blocked configurations exist since individuals can always be relocated. 
For this reason, the system can enter a stationary regime in which the
densities of unhappy individuals $u(t)$ remain constant on average.
We are thus interested in understanding the asymptotic properties of the dynamics, and characterizing the evolution to the stationary state. We study these properties in $d=1,2$ as a function of the density of vacancies $\rho_{0}$,
assuming $m=0$. \\
In $d=1$, for any $\rho_{0}$, $u(t)$ has an initial decay followed by a stationary state with asymptotic value $u_{\infty}(\rho_{0})$. 
The initial decay is exponential for large densities of vacancies ($\rho_{0} \to 1$), while it becomes power-law in the limit of vanishing $\rho_{0}$. 
The density $n(t)$ of interfaces between domains of unlike spins exhibits a similar behavior. Fig.~\ref{fig2} shows that for $\rho_{0} \ll 1$ the interface density follows the scaling law $n(t) \sim \rho_{0} \psi(\rho_{0} t^{1/2})$ with $\psi(x) \sim x^{-1}$ for $x \to 0$ and $\psi(x) \sim const$ for $x \to \infty$.  A similar power-law behavior ($u(t) \sim t^{-3/2}$) is found for $u(t)$ in this limit, but $u(\rho_0,t)$ does not satisfy any scaling law.\\
The behavior of the stationary densities as functions of $\rho_0$ is reported in the inset of Fig.~\ref{fig2}.
Both $u_{\infty}$ and $n_{\infty}$ have a maximum at some value $0 < \rho_{0}^{*} <1$, then vanish in the limits
$\rho_{0} \to 0 , 1$. In particular, for $\rho_{0} \to 1$, $u_{\infty}(\rho_{0})$ vanishes
as ${(1-\rho_{0})}^{3}$, while in the opposite limit  $n_{\infty} \sim \rho_0$ and $u_{\infty} \sim \rho_{0}^2$.\\
An explanation for the observed behavior of $u_\infty$ for $\rho_0 \approx 1$ can be given in terms of a phenomenological dynamics for $u(t)$. This takes the form $\dot u=-au+b$ where $a$ and $b$ describe processes which annihilate or create unhappy sites, respectively.  For $\rho_0\approx 1$ an unhappy individual with high probability becomes happy, once it is displaced, hence $a\simeq 1$. Unhappy individuals are created in processes such as $\ldots +--\ldots\rightarrow \ldots +-0\ldots$ where a displaced site leaves one of his neighbors in an uncomfortable neighborhood. For $\rho_0\simeq 1$, the leading contribution is given by initial configurations of three occupied sites, whose probability is $4[(1-\rho_0)/2]^3$, where the factor $4$ accounts for the degeneracy of the possible types. Hence $u(t)$ decays exponentially to $u_\infty=b/a\simeq (1-\rho_0)^3/2$, which agrees very well with numerical simulations.

A more elaborate analysis is instead necessary in order to explain the observed phenomenology in the limit $\rho_{0} \to 0$. It is convenient to think in terms of interfaces between clusters of opposite spin values. These can be thought of as 
particles, like in diffusion-limited annihilation processes.
For two adjacent clusters of opposite spins, the leading process is
the diffusion of interfaces, e.g. $ ++--- \to ++0-- \to +++--$. 
The other relevant processes are the creation and annihilation of clusters of size one, i.e. unhappy spins. The annihilation rate of unhappy spins is obviously $\propto n^2$, whereas the creation of interfaces involves ``vacancy mediated'' 
processes of the type  $ ++--- \to ++-0-$. Here the displacement of an individual makes one of his neighbors unhappy, and the latter can in turn move in the bulk of a cluster of individuals of the other species.
In the stationary state, the creation rate is proportional to the probability to find an empty site close to an interface, which is $\rho_{0} n$. 
The balance condition $\rho_{0} n \approx n^2$ implies that the stationary interface density is $n_{\infty} \propto \rho_{0}$ and hence that the creation rate is proportional to $\rho_0^2$.

The solution to a problem of diffusion-limited annihilation with input of particle pairs is well known \cite{racz} and it leads to the following scaling form for the density of interfaces, $n(\rho_{0},t) \sim \rho_{0}^{1/\delta} \psi(\rho_{0} t^{1/\Delta})$, with $\psi(x) \to const$ for $x \to \infty$ and $\psi(x) \to x^{-1/\delta}$ for $x\to 0$. 
The injection rate $\rho_{0}^2$ implies $\delta=1$ and $\Delta=2$, in perfect agreement with numerical simulations. At $\rho_{0}=0^+$ (single vacancy limit), the dynamics is purely diffusive, thus $n(\rho_{0}=0^+,t) \sim t^{-1/z}$, with $z=\delta \Delta =2$. In other words, the $\rho_{0} \to 0$ limit can be considered as a critical point for the one-dimensional unconstrained Schelling model. In this respect, the Schelling model for $\rho_0>0$ produces a dynamics with the same universal features of finite temperature coarsening dynamics in Ising like models with non-conserved order parameter \cite{racz}.

Contrary to $n(\rho_0,t)$, the density of unhappy individuals does not follow a scaling form.  Indeed one can write $u(\rho_0,t)=-\partial n(\rho_{0}=0,t) / \partial t +c\rho_0n$, with $c>0$ a constant. The first term is the probability to have a cluster of size one in the bulk of larger domains, whereas the second arises from the creation of unhappy individuals close to interfaces (see above). For early times ($t\ll \rho_0^{-1}$, as suggested by dimensional analysis) the first term dominates, hence $u(\rho_0,t)\sim t^{-3/2}$, whereas $u(\rho_0,t)$ converges to $n_\infty^2(\rho_0)$ for $t\gg \rho_0^{-1}$. 

We now turn to discuss the two-dimensional unconstrained dynamics, that was already studied in Ref.~\cite{kirman}, where it was suggested that some coarsening phenomenon should take place. 
We remark that, in the dynamics, voids are expected to be randomly distributed in the system. If $\rho_0$ is too large, a cluster of void sites starts percolating throughout the system, preventing the growth of the clusters of spins
\cite{stauffer} beyond a finite size. In a $d=2$ square lattice with Moore neighborhood,
the percolation transition takes place at $\rho_{0} \simeq
0.407$ \cite{stauffer}. Since here voids can move we expect this value to be only a lower bound for the real transition, that takes place at some $0.45 < \hat{\rho}_{0} < 0.5$.
For $\rho_{0} \geq \hat{\rho}_{0}$ we find that clusters grow up to a finite size depending on $\rho_{0}$ but independent of the system size $N$ \cite{preparation}.
\begin{figure}[t]
\centerline{
\includegraphics*[width=0.4\textwidth]{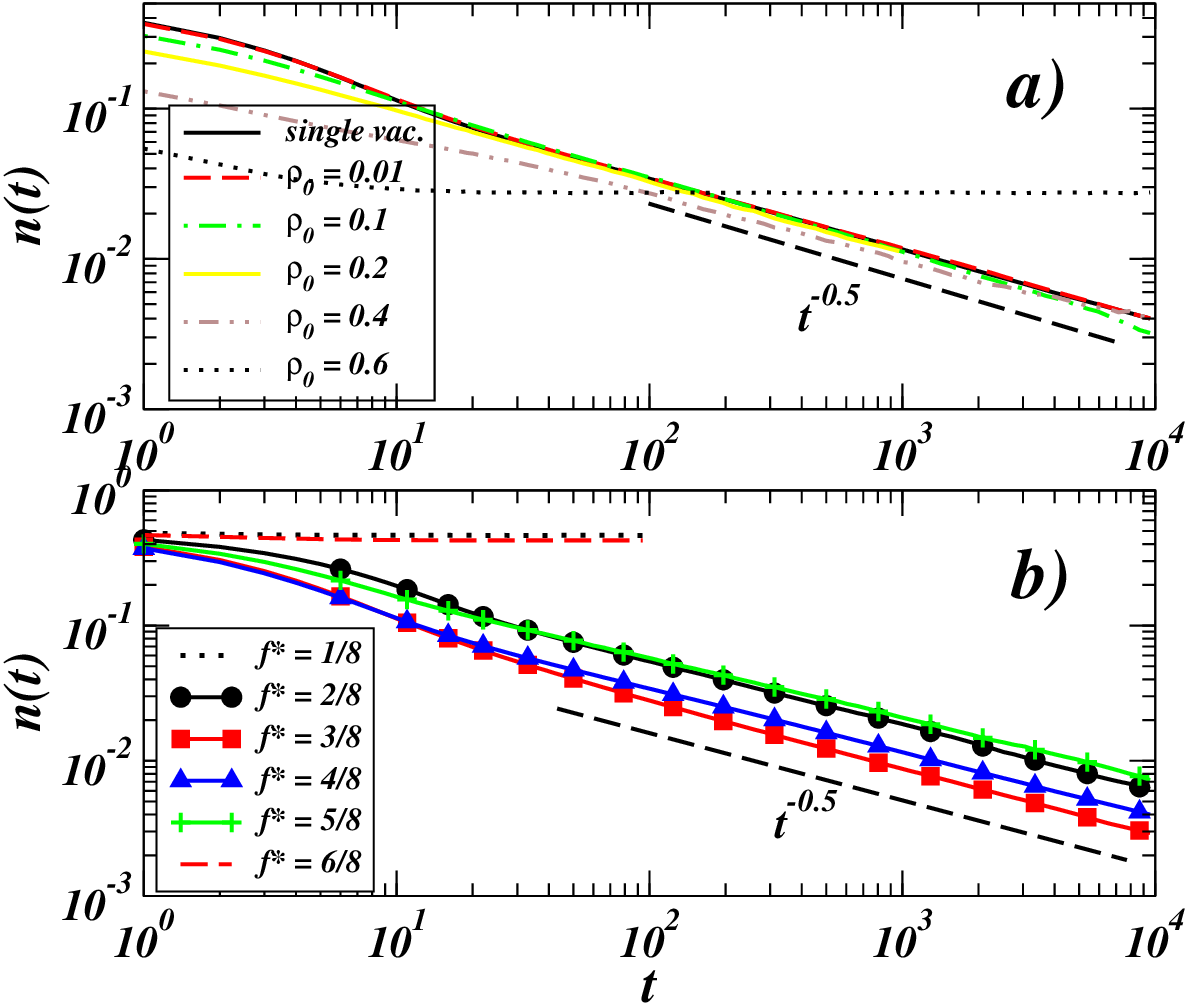}}
\caption{Density of interfaces $n(t)$ in a $2d$ system of size $L=10^3$: a) $f^{*}=1/2$, the coarsening process is present for $\rho_{0}<\hat{\rho}_{0}$; b) the systems does not order for some values of the threshold $f^{*}\neq 1/2$.
} 
\label{fig3}
\end{figure}
In contrast, below the void percolation transition, we observe convergence to a (quasi) ordered state, with two domains spanning the whole system.
A clear coarsening process with the typical length of clusters growing as $t^{1/z}$ with $z=2$ (see Figure~\ref{fig3}a) is observed for small $\rho_0$. Even if the magnetization is globally conserved, the value $z=2$ correctly describes a coarsening process with non-conserved order parameter in agreement with our findings in $1d$ and with renormalization group results for domain-growth scaling in the presence of long-range diffusion \cite{bray}.

Finally, it is interesting to discuss the behavior of the unconstrained model as a function of the maximal fraction $f^*$ of unlike neighbors which individuals tolerate (so far equal to $1/2$) in $d>1$. Figure~\ref{fig3}b shows that, for $\rho_0 \to 0$, the coarsening process is robust against the variation of $f^{*}$ in a wide range around $1/2$, as also found in Ref. \cite{kirman}. Segregation takes place even if individuals are satisfied with as few as 2 neighbors of their own type. If instead individuals are as tolerant as to be satisfied with just one like neighbors, then no coarsening and no segregation takes place and the system remains in a dynamic disordered state. No coarsening takes place also when individuals are extremely intolerant (see Fig. \ref{fig3}b). This is somewhat remarkable, because indeed a fully segregated state would be optimal, but it cannot be reached dynamically, as the system remains trapped in a disordered state.

{\em Conclusions -} In summary, we have investigated both analytically
and numerically the static and dynamic properties of the Schelling model
of segregation in one- and two-dimensional systems.
The constrained version of the model presents non-trivial static properties
characterized by the existence of a transition with symmetry breaking, whereas the unconstrained dynamics exhibit coarsening typical of systems with non-conserved order parameter. A further sharp transition takes place as the tolerance threshold ($f^*$) of individuals gets either very large or very small, with the system being trapped in disordered dynamical states.

\end{document}